# Fully connected entanglement-based quantum communication network without trusted node

Xu Liu[1], Rong Xue[1], Heqing Wang[2], Hao Li[2], Xue Feng[1,3], Fang Liu[1,3], Kaiyu Cui[1,3], Zhen Wang[2], Lixing You[2], Yidong Huang[1,3,4] and Wei Zhang[1,3,4, *]

[1]*Beijing National Research Center for Information Science and Technology (BNRist), Beijing Innovation Center for Future Chips, Department of Electronic Engineering, Tsinghua University, Beijing 100084, China*

[2]*State Key Laboratory of Functional Materials for Informatics, Shanghai Institute of Microsystem and Information Technology, Chinese Academy of Sciences, Shanghai 200050, China*

[3] *Frontier Science Center for Quantum Information, Beijing 100084, China*

[4]*Beijing Academy of Quantum Information Sciences, Beijing 100193, China*

*\* zwei@tsinghua.edu.cn*

Quantum communication is developed owing to the theoretically proven security of quantum mechanics, which may become the main technique in future information security. However, most studies and implementations are limited to two or several parties. Herein, we propose a fully connected quantum communication network without a trusted node for a large number of users. Using flexible wavelength demultiplex/multiplex and space multiplex technologies, 40 users are fully connected simultaneously without a trusted node by a broadband energy-time entangled photon pair source. This network architecture may be widely deployed in real scenarios such as companies, schools, and communities owing to its simplicity, scalability, and high efficiency.

**Introduction**

Quantum key distribution (QKD) has been regarded as a mature technique in quantum communication in security applications[1,2]. Since the first QKD protocol BB84[3] was proposed, QKD has been actively developed both in security proofing[4,5] and practical implementation[6,7]. The decoy state QKD[8-10] was proposed to solve the impurity of a single photon source to avoid photon number splitting. Subsequently, a measurement-device-independent QKD[11,12] was proposed, which is robust against attacks from the measurement device. In the last two years, twin-field QKD (TF-QKD)[13-15] was proposed based on single-photon interference, which can provide high key rates over long distances to surpass the rate-distance limit of repeaterless QKD. Based on TF-QKD, the secure distance of QKD was extended to 509 km experimentally[16].

However, an optimal method to build quantum communication networks based on QKD has yet to be developed. Quantum repeater-based networks[17-19] can be the ultimate blue print for constructing the global quantum Internet. However, quantum memory[20,21] and entanglement swapping[22,23] technologies remain to be improved for practical applications. Meanwhile, trusted node networks[24-27] have been developed and implemented widely. Trusted node networks are suitable for constructing a long distance backbone core network; however, they are inefficient for constructing multiple-user group networks. Furthermore, the security is compromised because every connected node in the network must be trusted, which is difficult to guarantee. Another type of quantum network based on active switches[28-30] has been developed. However, only some of the pairing users can be connected at a time. The network efficiency is limited to some duty cycles of switches. Moreover, additional time is required to reinitialize the new communication channel when the topology is changed[31]. In addition, a point-to-multipoint quantum network based on a passive beam splitter based on a single-photon point-to-point QKD[32,33] has been proposed, in which single photons from a central node are distributed to multiple users by a passive beam splitter. Every user must exchange keys with the central node, implying that the central node must be trusted.

The last type of quantum network is the fully connected quantum network without a trusted node, which can be based on the entanglement distribution among users. Every user can be connected directly to each other. A type of fully connected quantum network with four users based on wavelength multiplexing has been reported in a pioneering study[34]. To fully connect the four users, 12 wavelength channels are required. Namely, a minimum of $N \times (N-1)$ wavelength channels are required to fully connect $N$ users, which hinders the scheme to be scaled to a large user number. Furthermore, an improved scheme was proposed by introducing a $1 \times 2$ beam splitter[35]. The scheme supported an eight-user fully connected quantum network with 16 wavelength channels. However, it can be further improved. Recently, another type of fully connected quantum network was proposed[36]. In this scheme, resources of entangled photon pairs occupied with two correlated wavelength channels were directly distributed to eight users by a passive beam splitter to construct a fully connected subnet. To expand the user scale of the network, 16 such subnets were constructed using resources with different wavelength channel pairs. However, the connections between the subnets relied on a trusted central node.

In this study, we investigated a 40-user fully connected quantum communication network supported by a broadband energy-time entangled photon pair source, in which each user can simultaneously generate secure keys with every other user via a QKD. Five subnets were constructed using space multiplexing technology based on passive beam splitters. In each subnet, the entangled photon pairs with a correlated wavelength channel pair were randomly distributed to eight users, realizing a fully connected subnet. Furthermore, 10 additional correlated wavelength channel pairs of resources of entangled photon pairs were demultiplexed, and photons in these channels were flexibly multiplexed and distributed to different subnets, establishing connections between the subnets. Hence, the 40 users in the quantum communication network were fully connected without the assistance of trusted nodes. To the best of our knowledge, this is the largest experimentally demonstrated fully connected quantum communication network supported by a single quantum light source.

**Results**

**Network Architecture**

Based on composite space multiplexing and wavelength multiplexing technologies for the resources of entangled photon pairs, we proposed a network architecture that supports a fully connected quantum communication network. The signal and idler photons of the energy-time entangled state from a broadband quantum light source were distributed to all users in the network. Entanglement resources of 15 correlated wavelength channel pairs from a broadband quantum light source were required to fully connect the 40 users. An illustration of the network architecture is shown in Figure 1. The two wavelength channels whose subscripts are of opposite numbers belong to a specific correlated wavelength channel pair from the quantum light source, i.e., $(\lambda_1,\lambda_{-1})$, $(\lambda_2,\lambda_{-2})$, etc. To understand the network architecture more effectively, the network architecture can be segmented into two layers. A sketch of the first layer is shown in Figure 1(a). In this layer, photons with a specific pair of correlated wavelengths are distributed to $N$ users by a passive beam splitter. The signal and idler photons of this entanglement resource are randomly distributed to any user. Hence, each user will have coincidence events with any other user, thereby forming a subnet with a fully connected topology. If the user number of this subnet is appropriate, most photon pairs will be randomly distributed to two different users, which is a simple yet efficient approach to realize a fully connected network. It is clear that the rates of coincidence events between the users will decrease when the spitting ratio of the beam splitter increases. Hence, in this study, a high dimension time encoding method based on dispersive optics QKD was used for key generation between users to improve the utilization of coincidence events by generating multiple bits per coincidence.

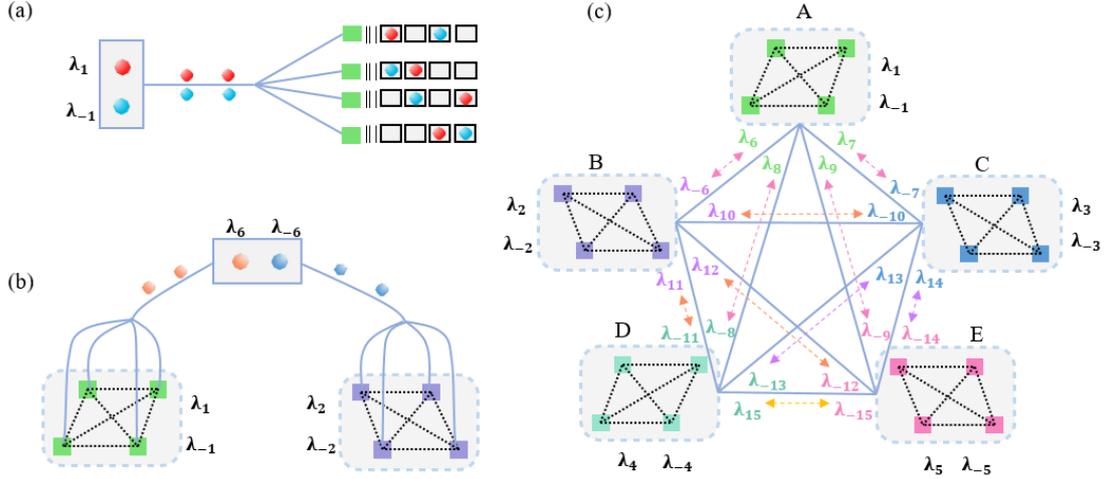

Figure 1 Illustration of network architecture depicting distribution of photon pairs of entanglement resources with different correlated wavelength channel pairs to users. Subscripts with opposite numbers represent wavelength channels corresponding to a specific entanglement resource. (a) Sketch of first layer. Entangled photon pairs with two correlated wavelength channels were distributed randomly to users by a passive beam splitter. As the signal and idler photons in a pair move randomly to the users, each user will have coincidence events with any other user. Hence, it constructs a fully connected quantum network without a trusted node, thereby forming a subnet in the network architecture. (b) Sketch of second layer. Green and purple squares denote users in two subnets, which were supported by two entanglement resources ($\lambda_1/\lambda_{-1}$ and $\lambda_2/\lambda_{-2}$) separately. To establish connections between these two subnets, an additional entanglement resource ($\lambda_6/\lambda_{-6}$) was used. The signal and idler photons of this resource were separated by wavelength division multiplexing components and distributed to users in these two subnets by the same passive beam splitters along with photons of wavelength λ1/λ-1 or λ2/λ-2 in the first layer. Hence, each user in one subnet will have coincidence events with any user in the other subnet by sharing an entanglement resource. Therefore, all users in the network were fully connected. (c) Based on this two-layer network architecture, a fully connected quantum network with five subnets was constructed in this study. Each user received six wavelength channels. Two of them corresponded to a pair of correlated wavelength channels for supporting the connection of users in the subnet. Four wavelength channels among the corresponding wavelength channel pairs were used to connect the users between different subnets.

A sketch of the second layer is shown in Figure 1(b). Two subnets are illustrated as two fully connected mesh graphs, which were supported by two independent entanglement resources with different correlated wavelength channel pairs ($\lambda_1/\lambda_{-1}$ and $\lambda_2/\lambda_{-2}$) from the quantum light source. An additional entanglement resource with the correlated wavelength channel pair of ($\lambda_6/\lambda_{-6}$) was introduced to connect the two subnets. The signal and idler photons with the entanglement resource of ($\lambda_6/\lambda_{-6}$) were separated by wavelength division multiplexing components and distributed to the two corresponding subnets. The photons were randomly distributed to the users by the same passive beam splitter along with photons of wavelengths ($\lambda_1/\lambda_{-1}$) or ($\lambda_2/\lambda_{-2}$) in each subnet. Therefore, each user in one subnet can have coincidence events with any user in the other subnet because of the correlation properties of entangled photon pairs with correlated wavelengths $\lambda_6$ and $\lambda_{-6}$. Based on the two-layer network architecture, any two users in the network can be fully connected without a trusted node. In this study, we realized a large-scale quantum communication network with 40 users based on this architecture, as shown in Figure 1(c). Five fully connected subnets (A, B, C, D, and E) were supported by five entanglement resources (from $\lambda_1/\lambda_{-1}$ to $\lambda_5/\lambda_{-5}$). Ten additional entanglement resources (from $\lambda_6/\lambda_{-6}$ to $\lambda_{15}/\lambda_{-15}$) were introduced to realize the full connections among the five subnets. Hence, each user in the five subnets can have coincidence events with any other user in the network, namely, every pair of users can share an entangled resource. The detailed wavelength allocation of users is given in Supplementary Materials (See Supplementary Table 1). For each user, only one fiber was connected to the entanglement resource provider. Through this fiber, the user received six wavelength channels. Two of them were the correlated wavelength channels from the entanglement resource supporting the connections of users in the subnet. The other four wavelength channels from the wavelength channel pairs of the entanglement resource were used to connect the users between different subnets.

**Experimental Setup**

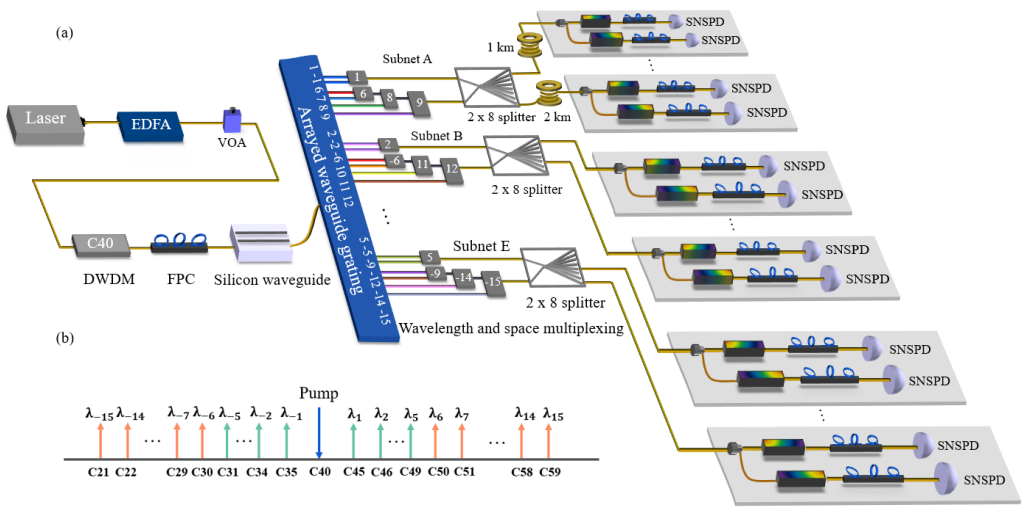

Figure 2 Experimental system of 40-user fully connected quantum network. (a) Energy–time entangled photon pairs were generated through spontaneous four wave mixing (SFWM) based on a silicon waveguide. Arrayed waveguide grating was used to demultiplex the entangled photon pairs

by wavelengths into 100-GHz-spaced ITU channels. After demultiplexing, the corresponding combinations of wavelength channels were multiplexed together by 100 GHz dense wavelength division multiplexing (DWDM) components and distributed to users in different subnets. The numbers on the DWDMs represent the subscripts of the corresponding wavelength channels. Each subnet received six wavelength channels. These photons were randomly distributed to the users by a passive beam splitter. In each user, normal and anomalous dispersion components were placed at two paths, followed by two NbN superconducting nanowire single-photon detectors (SNSPDs). The arrival time of photons can be used for key generation and security test based on symmetric dispersive optics QKD. (b) Mono-color pump light of the quantum light source was set at 1545.32 nm, which is the central wavelength of ITU channel C40. Generated signal and idler photons with correlated wavelength channels were symmetrically distributed around the central frequency of pump light. Subscripts of wavelengths with opposite numbers correspond to a correlated wavelength channel pair for a specific entanglement resource; corresponding ITU channels are indicated below wavelength symbols. Five correlated wavelength channel pairs (shown in green) were used to support five subnets. The other 10 correlated wavelength channel pairs (shown in orange) were used to connect users between subnets.

The experimental system of the 40-user fully connected quantum communication network without a trusted node is shown in Figure 2. In the experiments, broadband energy-time entangled photon pairs were generated by spontaneous four-wave mixing (SFWM) under continuous wave pumping in a silicon waveguide of length 3 mm. The central wavelength of the pump light was 1545.32 nm, corresponding to the International Telecommunication Union (ITU) channel of C40. Owing to the energy conservation of the SWFM process, the signal and idler photons were distributed symmetrically around the pump light wavelength. They were separated by an arrayed waveguide grating system based on their wavelengths with 100 GHz spacing (See Supplementary Fig. 4(a) in Supplementary Materials). Fifteen entanglement resources were extracted from the quantum light source, which corresponded to correlated wavelength channel pairs of C35/C45, C34/C46, …, C21/C59, as shown in Figure 2(b). The corresponding coincidences of entangled photon pairs with correlated wavelength channels can be seen in Supplementary Materials, Supplementary Fig. 4(b). The first five entanglement resources (represented in green) were used to support the connection of users in the five subnets. The remaining 10 entanglement resources (represented in orange) were used to connect users between the subnets. Subsequently, these wavelength channels were multiplexed by commercial dense wavelength division multiplexing components, as illustrated in Figure 1(c), and then sent to the passive beam splitters. In each subnet, the passive beam splitter distributed the input photons to all users randomly. The quantum light source, wavelength demultiplex/multiplex components, and passive beam splitters can be treated as a provider of entanglement resources for the network. Two specific users in subnet A received the photons from the provider through transmission of optical fibers of 1 and 2 km, separately. Other users connected to the provider using short fiber patch cords.

In each user, a normal dispersion component, an anomalous dispersion component, and two NbN superconducting nanowire single-photon detectors (SNSPDs) were equipped for performing symmetric dispersive optics QKD (DO-QKD)[36]. The symmetric DO-QKD was modified from the conventional DO-QKD scheme[37-39] to fully adapt to the entanglement distribution network based on passive beam splitters. High-dimensional encoding based on the time of recorded single photon detection events can be used in symmetric DO-QKD to improve the utilization of coincidence events by multi-bit key generation per coincidence.

**Experimental results**

First, the properties of the entanglement distribution were measured to verify the feasibility of the network architecture and to evaluate the quality of coincidences between the users. For each user, the photons were directly detected by the SNSPD. The results are shown in Figure 3. Figure 3(a) shows the typical results for two specific users in the same subnet. The five peaks show the results of coincidence counts in the five subnets (A, B, C, D, and E), which were supported by the resources of correlated wavelength channel pairs of (C35, C45), (C34, C46), (C33, C47), (C32, C48), and (C31, C49), respectively. For clarity, the coincidence peaks of the five subnets were plotted in the same figure with different offsets in the time delays. The time window for the coincidence measurement was 128 ps. It was observed that the coincidence to accidental coincidence ratios (CARs) of all the peaks were higher than 70. Figure 3(b) shows the typical coincidence results between users of different subnets. In each subnet, one specific user was selected to perform the coincidence measurement. Hence, 10 connections existed among the five users in different subnets. The 10 peaks in the figure show the coincidence results of the 10 connections with different time offsets for clarity. The time window for the coincidence measurement was 128 ps. The first coincidence peak marked as AB, which indicates the result for the users from subnets A and B, was supported by the entanglement resource with correlated wavelength channel pair ($\lambda_6$, $\lambda_{-6}$), and so on for the other coincidence peaks. All the peaks showed CARs exceeding 60. The results in Figure 3 show that the photon pairs distributed to any two users can be well discriminated under a narrow coincidence window by the coincidences, regardless of whether they are in the same or different subnets.

It is noteworthy that the average coincidence counts between two users in different subnets were smaller than those between two users in the same subnet. This is because the signal and idler photons were separately distributed to the users in different subnets. For the connection of two users in the same subnet, either of the signal and idler photons will reach each one of the users, thereby resulting in an almost two-fold coincidence count. The differences in coincidence counts shown in Figure 3(b) are primarily due to the differences in insertion loss induced by the wavelength division multiplexers for the photon pairs of different entanglement resources.

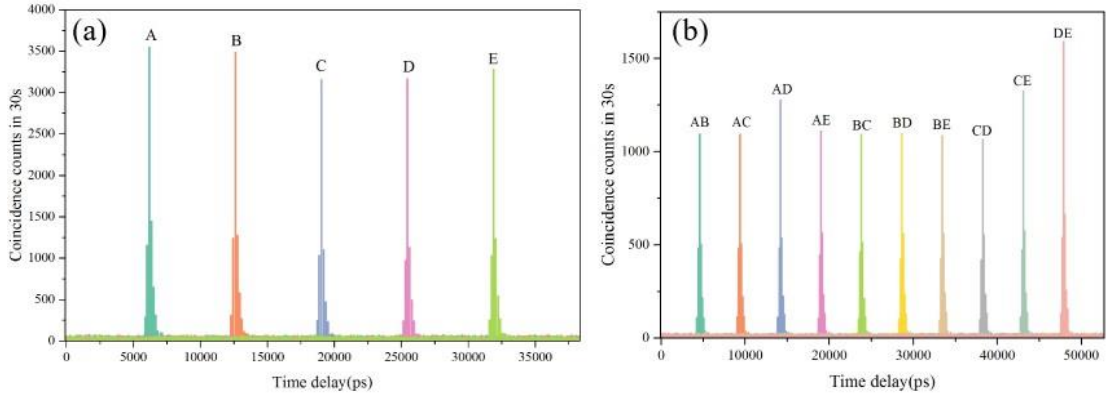

Figure 3 Experimental results of coincidence measurements between two typical users (a) in the same subnet, and (b) in different subnets. The coincidence time window for the coincidence measurements was 128 ps.

Subsequently, the performance of the QKD in this quantum communication network was measured using the setup shown in Figure 2(a). A symmetric DO-QKD was applied to realize secure key generation in all the links in the network. In this QKD scheme, the arrival time of photons was recorded and used for key generation and security tests. A high-dimensional time encoding process with three levels was optimized to attain the maximum secure key generation rates. Details of the symmetric DO-QKD are introduced in the Supplementary Materials. Figure 4 shows the measurement results. First, the secure key rates between any two users in subnet A were measured, as shown in Figure 4 (a). To form all the links in the subnet, 28 user combinations labeled by numbers along the x-axis existed. Because two users received photons through transmission fibers of 1 and 2 km, separately, these links had different transmission conditions. Link 1 had transmission fibers of 1 and 2 km on two sides, separately. Links 2–7 only had transmission fibers of 1 km on one side, whereas links 8–13 only had transmission fibers of 2 km on one side. Links 14–28 did not contain these transmission fibers. It was discovered that all the links exhibited similar performances. The average secure key rate was ~51 bps. This indicates that the fiber transmissions on the scale of local networks or campus networks did not affect the performance of the symmetric DO-QKD in this network. To demonstrate the secure key generation between two users in different subnets, we selected one user in each of the five subnets. Ten links existed among the five users. The performances of the symmetric DO-QKD of these links were measured, and the results are shown in Figure 4(b). The letters on the top of each result indicate the two subnets of the two users of the corresponding link. All the links with the user from subnet A (1–4) had transmission fibers of 1 km on one side. Other links (5–10) did not contain transmission fibers. Furthermore, it was observed that all these links demonstrated similar performances. The average secure key rate was ~22 bps, which was lower than that shown in Figure 4(a). This was consistent with the expectation explained in the discussion of the coincidence results of entanglement distribution. The corresponding results of QBER between the users in the subnet A and in each two of five subnets can be seen in Supplementary Materials, Supplementary Fig. 5. The QBER are all bounding at less than 8% by the bin sifting process.

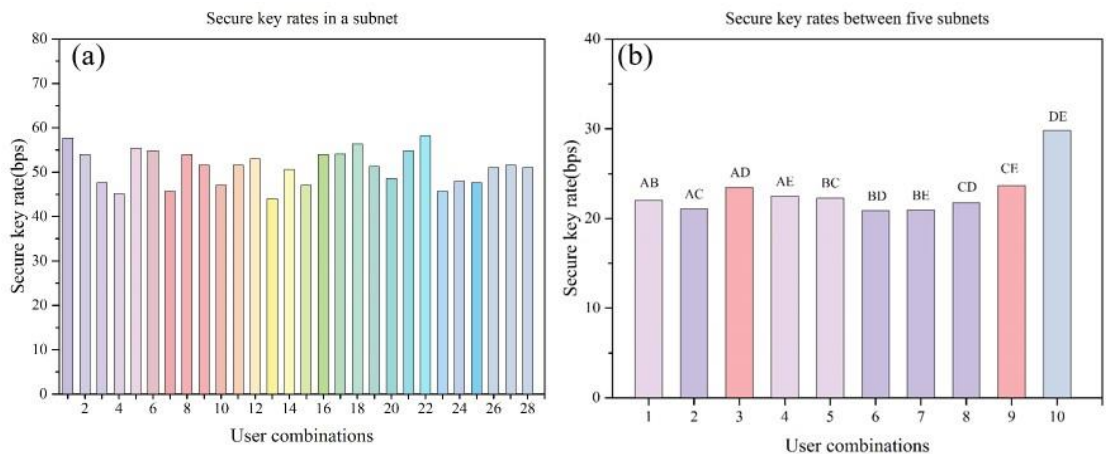

Figure 4 Performances of symmetric DO-QKD in the network. (a) Measured secure key rates between any two users in subnet A; 28 links labeled by numbers on x-axis exist. Link 1 comprised transmission fibers of 1 and 2 km at two sides, separately. Links 2–7 only contained transmission fibers of 1 km at one side, and links 8–13 only contained transmission fibers of 2 km at one side. Links 14–28 did not contain these transmission fibers. (b) Measured secure key rates between users in different subnets. Ten links existed among five users in five different subnets. The letters on top of each result indicate two subnets of two users in the corresponding link. All the links with the user from subnet A (1–4) contained transmission fibers of 1 km at one side. Other links (5–10) did not contain transmission fibers.

**Discussion**

In this study, we successfully demonstrated a 40-user fully connected quantum communication network without a trusted node, verifying the feasibility of the network architecture shown in Figure 1. Fifteen entanglement resources with different correlated wavelength channel pairs were provided by a broadband quantum light source. Based on multiport beam splitters, five eight-user subnets with fully connected mesh topologies were constructed using five entanglement resources. Ten other entanglement resources were used to connect the users between different subnets by flexible wavelength demultiplexing and multiplexing. Hence, the entanglement resources of the quantum light sources were distributed to the 40 users, realizing a fully connected entanglement distribution network. Any two users in the network had coincidence events owing to the entanglement resource shared by them; the events were then used to generate secure keys by a symmetric DO-QKD.

This architecture provides a simple approach for realizing large fully connected quantum communication networks without a trusted node. Only one quantum light source was required, and it should provide multiple entanglement sources by wavelength division multiplexing. The quantum light source can be realized using several methods, such as spontaneous parametric down conversion in PPLN crystals[34], waveguides[40], PPKTP[41], poled fiber[42,43], and SFWM in silicon waveguides[36]. The bandwidth of the quantum light source determines the quantity of entanglement resources that can be provided, and hence the number of users that can be supported by this network architecture. Meanwhile, in the network comprising 40 users, a user shares the entanglement resources with all the other 39 users and establishes a symmetric DO-QKD with them. However, only one set of symmetric DO-QKD setup is required for a user, and it is shared by all the 39 QKD links. Hence, although 780 QKD links exist in the network among the 40 users, only 40 QKD setups must be installed for each user. This characteristic significantly reduces the complexity and cost of establishing such a fully connected quantum communication network with multiple users. Combined with other active and passive routing technologies such as optical switch arrays[44], large-scale networks with more complicated topological structures can be supported. It is noteworthy that the basis of this quantum communication network is simple yet effective for realizing a multiple-user quantum entanglement distribution, which is the foundation of quantum Internet. Other entanglement-based quantum communication functions, such as quantum teleportation and entanglement swapping, are expected to be applicable to this network architecture.

The measured secure key rates in this network were approximately several tens of bits per second between the users, whether in the same or different subnets. Better QKD performances can be achieved if the performances of the quantum light source and single-photon detectors are further improved. The photon pair generation rate of the quantum light source can be improved by reducing the coupling loss of the silicon waveguide and enhancing the pumping level. More entanglement resources can be provided by utilizing more correlated wavelength channel pairs. Finally, the counting rates of the coincidence events can be improved significantly using single-photon detectors with better detection efficiencies and shorter recovery times (limiting the maximum counting rate of the detectors). Higher secure key rates and larger network scales can be achieved in the fiber transmission distance or user number by the performance improvements.

**Methods**

In the experiments, the entanglement resources of fifteen correlated wavelength channel pairs are distributed to forty users in the network by flexible wavelength demultiplexing, multiplexing and space multiplex technologies. Any two users in the network are connected simultaneously by sharing a paring of entangled photons. The entanglement resources are provided by a broadband entangled photon pair source based on a silicon waveguide with a length of 3 mm. Thanks to the high nonlinearity and the low dispersion property of the silicon waveguide, broadband photon pairs with energy-time entanglement are generated through spontaneous four wave mixing (SFWM) under CW pumping. The signal and idler photons of entangled photon pairs are generated symmetrically with the central frequency of pump light. Then the entangled photon pairs are divided according to their wavelengths by an arrayed waveguide grating (AWG) filtering system (5.5 dB) in which each output port of AWG are cascaded with a dense wavelength division multiplexer (DWDM) with corresponding matched wavelength. After this filtering system, the spectra of entangled photon pairs are split into fifteen pairs of wavelength channels, namely 30 channels. And the signal to noise ratio of entangled photons of different channels can be achieved to ~120 dB. The filtering DWDM channels have a ~0.8 nm (100 GHz) bandwidth. After this wavelength de-multiplexing process, the corresponding wavelength combinations shown in Figure 1(c) in the manuscript and Supplementary Table 1 are multiplexed by flexible WDM system shown in Figure 2. In each combination set, two of them are correlated in wavelength channel, which are for the connection of users in the subnet. Other four wavelength channels are used to connect the users between different subnets. The wavelength combination sets are then distributed to the different subnets by multi-port beam splitters (10.4 dB). Each WDM component has an insertion loss of 0.5 dB. All these parts discussed can be treated as an entanglement resource provider. The entanglement resources are distributed to the users in the network. At each user, a normal/an anomalous dispersion component (DCM, Proximion AB, Sweden) and two NbN superconducting nanowire single-photon detectors (SNSPD, fabricated by SIMIT, CAS, China) are equipped for performing the symmetric DO-QKD. The group velocity dispersions of dispersion components are ±1980ps/nm, whose insertion loss is ~3dB. The SNSPDs have a detection efficiency of ~70% with a dark count rate of ~100. The fiber polarization controllers before the SNSPD are used to maximize the detection efficiencies. In the experiments, 1 km and 2 km transmission fibers are deployed to two users in the subnet A. Hence, the QKD between different users are performed with varying link lengths and specifications. The transmission fibers are all single mode at telecom band of 1550 nm.

**Data availability**.

The data supporting the results of this study are available from the corresponding author upon request.

**Acknowledgements**

We acknowledge the support of National Key R&D Program of China (2017YFA0303704, 2018YFB2200400, 2017YFA0304000), National Natural Science Foundation of China (NSFC) (61875101, 91750206, 61575102, 61621064), Beijing National Science Foundation (BNSF) (Z180012), Beijing Academy of Quantum Information Sciences (Y18G26), and the Tsinghua University Initiative Scientific Research Program.

**Author contributions**

Wei Zhang and Xu Liu proposed the scheme and took the theoretical analysis. Xu Liu, Rong Xue and Wei Zhang performed experiments and analyzed data. Xu Liu and Wei

Zhang wrote the manuscript. Yidong Huang revised the manuscript and supervised the project. Xue Feng, Fang Liu, Kaiyu Cui contributed to discussion of this study and the revision of the manuscript. Heqing Wang, Hao Li, Zhen Wang and Lixing You provided superconducting nanowire single photon detectors and revised the manuscript.

**Additional information**

**Supplementary Information** accompanies this paper

**Competing interests:** The authors declare no competing interests.